\documentclass[12pt,
showkeys, rotate, epsfig]{article}
\usepackage{epsfig}
\usepackage{graphicx}
\usepackage{amssymb}
\usepackage{amsthm}
\usepackage{amsmath}
\usepackage{amssymb,amsfonts}

\newcommand{\bz}{\bar{z}}
\newcommand{\een}{\end{subequations}}
\newcommand{\ben}{\begin{subequations}}
\newcommand{\be}{\begin{eqnarray}}
\newcommand{\ee}{\end{eqnarray}}

\renewcommand{\d}{\partial}

\begin{document}
\centerline{ \hfill hep-th/0406025 } \vspace{3.5cm}

\begin{center}
{\Large \bf A Conical Tear Drop as a Vacuum-Energy Drain for \\
\vskip.05in
  the Solution of the Cosmological Constant Problem
}
\end{center}
\vspace{1.cm}

\centerline{
{\normalsize \bf Alex Kehagias
}
} \vspace{0.3cm}
\centerline{\rm \small Department of Physics, Ethniko Metsovio
Polytechnio,
 GR-15773, Zografou, Athens, Greece}

\vspace{.5cm} \centerline{\em e-mail : kehagias@mail.cern.ch}
\vspace{2.cm}

\centerline{\normalsize \bf Abstract}
\begin{quote}
\indent
We propose a partial solution to the cosmological constant problem
by using the simple observation that a three-brane  in a
six-dimensional bulk is flat.  A model is presented in which
Standard Model vacuum energy is always absorbed by the transverse
space. The latter is a tear-drop like space with a conical
singularity, which preserves bulk supersymmetry and gives rise to
conventional macroscopic 4D gravity with no cosmological constant.
Its cone acts like a drain depleting vacuum energy from the
three-brane to the tear drop increasing its volume. We stress that
although gravity is treated classically, Standard Model is handled
quantum-field theoretically and the model is robust against
Standard Model corrections and particular details.
 The price paid is the presence of boundaries which are nevertheless physically harmless by
appropriate boundary conditions.

\end{quote}
\vspace{1cm}

\newpage
After so many years of its introduction and so
intensive effort to solve it, the cosmological constant problem
remains one of the most  provoking and challenging problems in modern
physics. Many attempts have been initiated with partial or no
success as one may verify from the huge literature on the subject
(see \cite{Weinberg} for reviews and references).
Among the many proposals, the brane-world scenario, according to
which our world is a three-brane floating in a higher-dimensional
bulk, opened up the possibility of reconsidering  the cosmological
constant problem on a new basis~\cite{Arkani-Hamed:2000eg}--\cite{Schmidhuber:2000cm}
In this framework, we will present here a new approach to the
problem which is based on the simple observation that a p-brane,
independently of its tension, in a p+3-bulk is always flat
\cite{KR}. As a result, a three-brane in a six-dimensional
spacetime will always be flat for any value of its tension.
However, there exists a restriction: The compactness of the
transverse two-dimensional space, required for the existence of
conventional 4D gravity on the brane, is achieved for special
values of the brane tension, or otherwise, additional branes of
correlated tensions are needed
\cite{Chen:2000at}--\cite{Burgess:2004kd}, so that some kind
fine-tuning cannot be avoided \cite{Shap}. We bypass this
constraint by adopting a non-compact transverse 2D space of finite
volume in order to have a massless 4D graviton. Such spaces
proposed in the past~\cite{Rubakov:1983bz}--\cite{Cohen:1999ia}
and it is known
that their non-compactness creates problems, the most dangerous of which
are connected with conservation laws (like energy, charge e.t.c). These problems however, may be avoided
by appropriate boundary conditions and non-compact spaces of finite volume can be physically acceptable.

Here we employ as internal 2D space a finite-volume cone. It
may soak up all vacuum energies produced by fields confined in the three-brane at its apex
as long as the bulk cosmological constant is zero.
We should stress that the cosmological constant we are refer to is
connected to the pure Standard Model (SM) sector (SM vacuum energy $\sim \Lambda^4$, for a cutoff
$\Lambda$).  We are ignoring quantum gravity
contributions (due to graviton loops in Feynman diagrams), which
although non-zero, are suppressed by inverse powers $1/M_*^2$ of the bulk
Planck mass $M_*$. We also mention that the three-brane preserves half of the bulk supersymmetries.

To make the discussion concrete,
let us consider a three-brane in a 6D  spacetime with coordinates ($x^\mu,y^m$),
$\mu,\nu=0,...,3,~~m,n=1,2$.
The 6D Einstein equations are \be R_{MN}-\frac{1}{2}
g_{MN}R=\frac{1}{M_*^4} T_{MN}\ ,\ \ \ M,N=0,...,5 , \label{einst}
\ee
where $M_*$ is the  6D Planck mass and the energy-momentum
tensor on the right-hand side of eq.(\ref{einst}) for the
three-brane configuration is given by \be T_{\mu\nu}= -\rho\
g_{\mu\nu}\ ,\ \ \ \ \ T_{mn}=0\ . \label{energy} \ee  In particular, if the brane
is localized at $y^m=0$ in the transverse space, its energy
density is of  the form $\rho=\rho_0 \  \delta^{(2)}({\bf y})$.
Einstein equations (\ref{einst}) can also be written in the form \be
R_{MN}=\frac{1}{M_*^4}\left( T_{MN}-\frac{1}{4}g_{MN} T\right)\,,
\ee
where $T=T_{MN}g^{MN}$ is the trace of the energy-momentum
tensor and using eq.~(\ref{energy}),
 we find that
\be R_{\mu\nu}=0
\ ,\ \ \ R_{mn}=\frac{1}{M_*^4}
\ \rho\
g_{mn}\, . \label{pp} \ee
As a result, a three-brane in a 6D bulk spacetime is flat irrespectively of any
vacuum energy produced by phase transitions, supersymmetry
breaking or quantum effects on the three-brane world-volume. All
these effects can only be seen on the transverse two-dimensional
space, and  will be absorbed into the curvature of the
two-dimensional geometry. It should be noted that this discussion is not restricted to three-branes
but it is rather general and one may easily show that the world-volume of  a codimension $2$ brane
is always Ricci-flat.
 This simple
observation, that the flatness of a three-brane in a
6D  bulk spacetime is automatic and independent of
any vacuum energy on the brane is our proposal for  the vanishing
of the cosmological constant in four dimensions.
However, there is an obstruction. Namely, the transverse space should be as usual a
compact two dimensional surface
since the conventional four-dimensional physics should be recovered
for length scales larger that the characteristic length (radius) of the compact surface.
In this case then, taking the trace of eq.(\ref{pp}) and integrating it over
the transverse 2D compact space we get
\be
4\pi \chi=2\frac{\rho_0}{M_*^4}\, ,  \label{eul}
\ee
where $\chi$ is the
Euler number.
For a compact space, $\chi$ should be $2-2g$ where $g=0,1,...$ is the genus of the 2D surface. As as result, compactness
of the 2D space spoils our attempt, since (\ref{eul})
is not valid for any value of $\rho_0$ but rather for some special value.
Alternatively, putting a three-brane at some point with some arbitrary $\rho_0$,
there should be  conical singularities somewhere, produced by other three-branes of fine-tuned
tension such that eq.(\ref{eul}) is satisfied. Definitely this is not a solution
to the cosmological constant problem as special values of the brane tension or fine-tunings are required.
A simple way out of this is to assume a non-compact transverse space as for example
surfaces with some points removed. In this case, the condition (\ref{eul})
is relaxed to
\be
2-2g+\sum_{i=1}^n(\frac{\delta_i}{2\pi}-2)=\frac{\rho_0}{2\pi M_*^4}  \label{eul2}
\ee
for $n$ points removed of deficit angle $\delta_i$ each, from the genus $g$ surface.
 For a sphere for example and one point removed, we
have actually that $\rho_0=M_*^4 \delta$.

Adopting the proposal of a non-compact internal space,
we are facing a new problem. A smooth non-compact space has
infinite volume so that the four-dimensional Planck mass
$M_{P}^2=M_*^4 V$ will be infinite. As a result, gravitational
interactions will be actually six-dimensional and not
four-dimensional as we would like. The solution here is to assume
that the non-compact surface has finite volume. In this case we
expect singularities and  several pathologies like continuous
spectra,  violation of conservation laws for  energy, momentum, angular momentum e.t.c.
caused by possible leakage from the singular points. Thus, in order our proposal to be viable, all these
pathologies should be avoided. Then, the three-brane will be
 always
flat for any value of its tension neutralizing this way any
four-dimensional cosmological constant,  at least the one which is connected to  vacuum energies.
We will
present below such a solution based on the tear-drop solution of
Gell-Mann and Zwiebach \cite{Gell-Mann1},\cite{Gell-Mann2}.

The set up of our proposal is the following. We consider a three-brane with all SM (or its extensions)
 degrees of freedom confined to it. The ambient space is a six-dimensional bulk spacetime with no cosmological constant.
 SM fields interact with the bulk through gravitational interactions. There are also two scalars, an  axion $b$ and a dilaton-like
$\phi$, which are combined into the complex scalar
$\tau=\tau_1+i\tau_2=b+ie^{-\phi}$. The latter parametrizes
$SL(2,\mathbb{R})/U(1)$ and it is  coupled to 6D gravity in the
presence of a three-brane. The action is
\be
S_6=\frac{1}{2}\int
d^6 x\, \sqrt{-g}\,
M_*^4\left(R-\frac{1}{2}\frac{\d\tau\d\bar{\tau}}{\tau_2^2}\right)+\int
d^2y \int d^4 x\, \sqrt{g}\, {\cal{L}}_{SM} \, \delta^{(2)}(y) \label{action}
\ee
where ${\cal{L}}_{SM}$ is the Lagrangian describing the SM  physics on
the three-brane at ${\bf y}=0$.
It should be stressed that we do not couple the dilaton to the three-brane. This is possible if the dilaton is part of a bulk
neutral hypermultiplet for example, and thus does not couple at all to the SM fields, or
it is fixed on the brane, if couples, i.e., $\delta S_{brane}/\delta \phi=0$ has a solution
$\phi={\rm const}$.
We may take into account all SM
quantum corrections by integrating  out  SM degrees of freedom and replace the
brane action by the full 1PI effective action $\Gamma_{\rm eff}^{SM}$.
 The  equations of motion are then
\be
&&R_{MN}=\frac{1}{4}\frac{\d_M \tau\d_N\bar{\tau}}{\tau_2^2}+\frac{1}{4}\frac{\d_N \tau\d_M\bar{\tau}}{\tau_2^2}+
\frac{1}{M_*^4}\left( T_{MN}-\frac{1}{4}g_{MN}T\right)\, ,\label{ein} \\
&&\nabla^2\tau-\frac{2\d_M\tau\d^M\tau}{\bar{\tau}-\tau}=0\, ,\label{sc}
\ee
where $T_{MN}=2\delta\Gamma_{\rm eff}^{SM}/\delta g_{MN} \delta({\bf y})$ is the brane energy-momentum
tensor.
In the vacuum we have $$\Gamma_{\rm eff}^{SM}=-\int d^4x \sqrt{g} V_{\rm eff}$$
where $V_{\rm eff}$ is the SM effective potential. As a result, the brane energy-momentum
is of the form eq.(\ref{energy}) with $\rho=\rho_0 \  \delta^{(2)}({\bf y})$ and
 $\rho_0$ is the value of the effective potential at its extremal value, $\rho_0=<\!\!V_{\rm eff}\!\!>$, i.e.,
the cosmological constant.
Assuming that
\be
ds_6^2=g_{\mu\nu}dx^\mu dx^\nu+ g_{mn}dy^m dy^n\, ,
\ee
and $\tau=\tau(y^m)$,  eq.(\ref{ein})
turns out to be
\be
&& R_{\mu\nu}=0\, , \\&&
 R_{mn}=\frac{1}{4}\frac{\d_m \tau\d_n\bar{\tau}}{\tau_2^2}+\frac{1}{4}\frac{\d_n \tau\d_m\bar{\tau}}{\tau_2^2}+
\frac{1}{M_*^4}\rho_0 \  \delta^{(2)}({\bf y})g_{mn} \label{ein2}
\ee
Thus, the four-dimensional space is always flat independently of the three-brane tension.
Writing the transverse space metric in complex coordinates
$(z,\bar{z})$, we have
\be
ds_6^2=\eta_{\mu\nu}dx^\mu dx^\nu+ e^{2A(z,\bar{z})}dz d\bar{z}\, , \label{MM}
\ee
and the scalar equation eq.(\ref{sc}) turns out to be
\be
\d\bar{\d}\tau-\frac{2\d\tau\bar{\d}\tau}{\tau-\bar{\tau}}=0\, . \label{eq1}
\ee
Eq.(\ref{eq1}) is solved for holomorphic (antiholomorphic) $\tau$, $\tau=\tau(z)$ ($\tau=\tau(\bar{z})$) and
the second equation in eq.(\ref{ein2}) is then written as
\be
-2\d\bar{\d}A=\frac{1}{4}\frac{\d\tau\bar{\d}\bar{\tau}}{\tau_2^2}+\frac{\rho_0}{M_*^4}\delta^{(2)}(z)\, . \label{14}
\ee
The solution to eq.(\ref{14}) is
\be
A=\ln\tau_2^{1/2}-\frac{1}{2\pi}\frac{\rho_0}{M_*^4}\ln \frac{|z|}{|z_0|} +f(z)+\bar{f}(\bar{z})
\ee
and the 2D metric turns out to be
\be
ds_2^2=|F(z)|^2 {\Big{|}\frac{z}{z_0}\Big{|}}^{-\frac{\rho_0}{\pi M_*^4}}\tau_2(z,\bar{z})dz\,d\bar{z} \, , \label{2dd}
\ee
where $F(z)=\exp{2f(z)}$. Note that, for any homolorphic $\tau=\tau(z)$, there exists a corresponding 2D metric (\ref{2dd}) and we will
assume that $\tau$ is regular at $z=0$ approaching a constant, as nothing special for the scalars happens at this point. Here we will consider
\be
\tau=i\frac{R^b+iz^b}{R^b-iz^b} \label{ttt},
\ee
which leads to the 2D metric
\be
ds_2^2={\Big{|}\frac{z}{z_0}\Big{|}}^{-\frac{\rho_0}{\pi M_*^4}}\left(1-\Big{|}\frac{z}{R}\Big{|}^{2b}\right)dz d\bar{z} \, . \label{2d}
\ee

Eq.(\ref{ttt}) is easily recognized as the conformal mapping of the upper-half plane to the  disc of radius $R$ in the
complex plane, whereas the Gell-Mann-Zwiebach case corresponds to $\rho_0=0,~ b=1$. Clearly, there exist a singularity at $|z|=R$ and a deficit angle
\be
\delta=2\pi a\, , ~~~a=\frac{\rho_0}{2\pi M_*^4} \label{aaa}
\ee
as we already know from eq.(\ref{eul2}). ( For a  metric like $|z^{-\delta/2\pi}dz|^2$ we may define $\zeta=z^{1-\delta/2\pi}$
such that the metric is the conventional $|d\zeta|^2$. However, as z undergoes a $2\pi$ rotation around zero, $\zeta$
does not complete a full circle but rather leaves a deficit $\delta$.)

This space is definitely non-compact and its volume is
\be
V_2=2 \pi \Big{|}\frac{z_0}{R}\Big{|}^{-2a}\frac{b\, R^2}{ (1-a)(1+b-a)}\, , ~~~~ ~~~\mbox{for} ~~~
a<1\, , ~~b>0
\ee
which is finite as the condition $a<1$ is easily satisfied for $\rho_0=M_s^4$ with  $M_s<M_*$, where $M_s$ is the 4D supersymmetry breaking scale.
\vspace{.5cm}

\noindent
 There is a number of points we would like now to discuss. These
 are consistency checks for the solution we propose and include,
 conservation of energy, momentum {\rm etc}.,  Newton's law on the
 three-brane, bulk supersymmetry, nearby curved solutions and the
 4D effective description and the Weinberg's theorem.
 \vspace{.3cm}

\noindent
 {I. \bf Conservation laws}
\vspace{.2cm}

 It is clear that the symmetries of the background are space-time translations and rotations and  U(1) rotation along the longitudinal (brane)
 and transverse directions, respectively. These symmetries are connected with conservations  laws and in particular with energy,
 momentum, angular-momentum and
 U(1) charge. However,
 these quantities may not be conserved due to the non-compactness of the transverse space since
 energy or momentum, for example, may leak from the boundary of the space.
 This situation is reminisent to open strings as the world-volume of the latter
 have a boundary. In that case, Neumann boundary conditions prevent momentum to flow off the ends of the string.
 Here, we hope that there are similarly  boundary conditions that make the $r=R$ boundary physically acceptable.
    To examine this, let us  consider
a massless scalar field $\Phi$ in the 6D spacetime which satisfied
\be
\frac{1}{\sqrt{-g}}\partial_M\left(\sqrt{-g}g^{MN}\d_N\Phi\right)=0\, .\label{scalar}
\ee
The spacetime metric around $r=R$ is of the form $\varrho^{-1} (d\varrho^2+d\varphi^2),$ where $\varrho=1-|z|/R$ and the solution to
eq.(\ref{scalar}) are Airy functions. What we would like to stress, is the existence of conserved quantities, like energy, momentum , angular
momentum and $U(1)$ charge which are connected to the corresponding Killing vectors
$\xi^{(a)\mu}= \delta^{a\mu}\, ,\xi^{(ab)\mu}=x^a \delta^{b\mu}-x^b\delta^{a\mu}$
and $\xi^m=\epsilon^{mn}y_n$ through the relation $J^{(a)M}=T^{MN} \xi^a_N$ where $T_{MN}$
is the energy-momentum
tensor of the massless scalar field $\Phi$. The conservation of the above quantities is expressed by $\nabla_M J^{(a)M}=0$, which, integrated over
all space-time demand the flux through the singular boundary to vanish, i.e.,
\be
\sqrt{-g}J^{(a) r}\Big{|}_{r=R}=0.
\ee
For momentum conservation for example, we get the Newmann boundary condition
\be
\frac{\partial \Phi}{\d r}\Big{|}_{r=R}=0\, ,
\ee
which is enough to prevent not only momentum but also  energy, angular-momentum and $U(1)$ charge to flow off the boundary.

Turning to metric perturbations
$g_{MN}=\bar{g}_{MN}+h_{MN}$, where $\bar{g}_{MN}$ is the background metric, it is known that
 for any isometry group $G$ generated by  the Killing vectors $\xi_M^A\, , A=1,...,\dim G$,
 there correspond  conserved charges \cite{Abbott:1981ff}
 \be
 Q^A=M^4_*\oint dS_p \sqrt{-\bar{g} } \left(\bar{D}_B K^{0pNB}-K^{0qNp}\bar{D}_q\right)\xi_N^A\, . \label{ch}
 \ee
 Here, $p,q=1,2,..5$ parametrize a fixed-time hypersurface and
 \be
K^{MNKL}=\frac{1}{4}\left( \bar{g}^{ML}\bar{h}^{NK}-
\bar{g}^{MK}\bar{h}^{NL}+\bar{g}^{NK}\bar{h}^{ML}-\bar{g}^{NL}\bar{h}^{MK}\right)\, ,
\ee
with $\bar{h}_{MN}=h_{MN}-\frac{1}{2}\bar{g}_{MN} h^K_K$.
In our case the boundary integral (\ref{ch}) has two pieces, one from integrating over the volume along the three-brane and the boundary of
the transverse space and one from integrating over the volume of the transverse space and the boundary along the brane directions. The
second part is regular as the volume of the transverse space is finite. We should examine only the first part, namely integration over
the boundary of the tranverse space which is just an $S^1$ parametrized with the angular variable $\varphi$.
To study energy conservation we should consider the  Killing vector $\xi_M=\delta^0_M$,
which generates time translations. The corresponding charge, the mass $M$, is
\be
M=2 b \,M^4_* |z_0|^{2 a}\, \varrho \int_0^{2\pi} d\varphi   \left(e^{+i\varphi }D_m K^{0\bar{z}0m}+e^{-i\varphi}D_mK^{0z0m}\right)
\ee
where $\varrho=1-|z|/R$ so that the boundary is now at $\varrho=0$.
This expression is exactly the same as in \cite{Gell-Mann2} and the result is that the  Neumann and Dirichlet boundary conditions
\be
\frac{dh_{\mu\nu}}{d r}\Big{|}_{r=R}=\frac{dh_{z\bar{z}}}{d r}\Big{|}_{r=R}=0\, , \\
h_{\mu z}\Big{|}_{r=R}=h_{z z}\Big{|}_{r=R}=h_{\mu \bar{z}}\Big{|}_{r=R}=h_{\bar{z}\bar{z}}\Big{|}_{r=R}=0
\ee
ensure not only the conservation of energy but the conservation of momentum, angular-momentum and $U(1)$ charge as well.
\vspace{.3cm}

\noindent
{\bf II.  KK modes} \vspace{.2cm}

Another danger is that although the 4D Planck mass is finite with a finite-volume internal space as in our case, it may happen that
the spectrum of KK modes is continuous with no mass gap due to the non-compactness of the internal space. Then
 gravity may be 6D with gravitational interactions of the form $1/r^3$ instead of the conventional
$1/r$ behavior.
In our case it is relatively easy to see that the spectrum of scalar excitations in the internal space have discrete spectrum. Indeed
by writing the 4D graviton $h_{\mu\nu}(x,z,\bar{z})=h_{\mu\nu}(x)\psi(z,\bar{z}),$ we find that the spectrum of the 4D graviton is related to the
eigenvalue problem $-\nabla^2\psi=M^2\psi$ on the transverse 2D space.
For the metric (\ref{2d}) and with $z=r e^{i\varphi}$
we get
\be
-\frac{1}{{r_{\mbox{\tiny 0}}^{2a}}}\frac{r^{2 a-1}}{\left(1-(r/R)^{2b}\right)}\left(\frac{\partial^2\psi}{\d r^2}+\frac{\partial\psi}{\d r}+\frac{1}{r}
\frac{\partial^2\psi}{\d
\varphi^2}\right)=M^2\psi
\ee
Expanding $\psi$ in Fourier modes, $\psi(r,\varphi)=\sum_{n=0}^{\infty} \psi_n(r)e^{i n\varphi}$,
the above equation
can be solved exactly for $b=1-a$. The regular solution at $r=0$ is
\be
\psi_n(r)=C_0\,  r^n  e^{-\frac{1}{2} \frac{M r_{ \mbox{\tiny 0}}^{a}}{b} \left(\frac{r^{2}}{R}\right)^b}
        {}_1F_1\left[\frac{1}{2}+\frac{n}{2b}-\frac{M R^{b}r_{\mbox{\tiny 0}}^{a}}{4b},1+\frac{n}{b},
        \frac{M r_{ \mbox{\tiny 0}}^{a}}{b} \left(\frac{r^{2}}{R}\right)^b\right]
         \ee
By imposing Neumann boundary conditions at $r=R$ we get the condition
\be
 {}_1F_1\left[\frac{3}{2}+\frac{n}{2b}-\frac{MR^{1-a}r_{\mbox{\tiny 0}}^{a}}{4b},2+\frac{n}{b},\frac{M R^b r_{\mbox{\tiny 0}}^{a}}{b}\right]=0\, ,
 \label{f1}
 \ee
This equation can be solved graphically as shown in Fig.1 for $n=0,1,2,3$. Clearly we have  a discrete spectrum with a massless mode $\psi_0=C_0$
corresponding to the 4D massless graviton. We thus have the conventional 4D $1/r$  Newton law, while the KK modes give rise to the Yukawa-type
corrections of strength 2 and range $\sim 1/V_2^{1/2}$~\cite{Kehagias:1999my} with the well-known constraints on $V_2$, the volume of the internal space
\cite{dim}.
\\
\vspace{.5cm}

\centerline{\epsfxsize=145pt\epsfysize=115pt\epsfbox{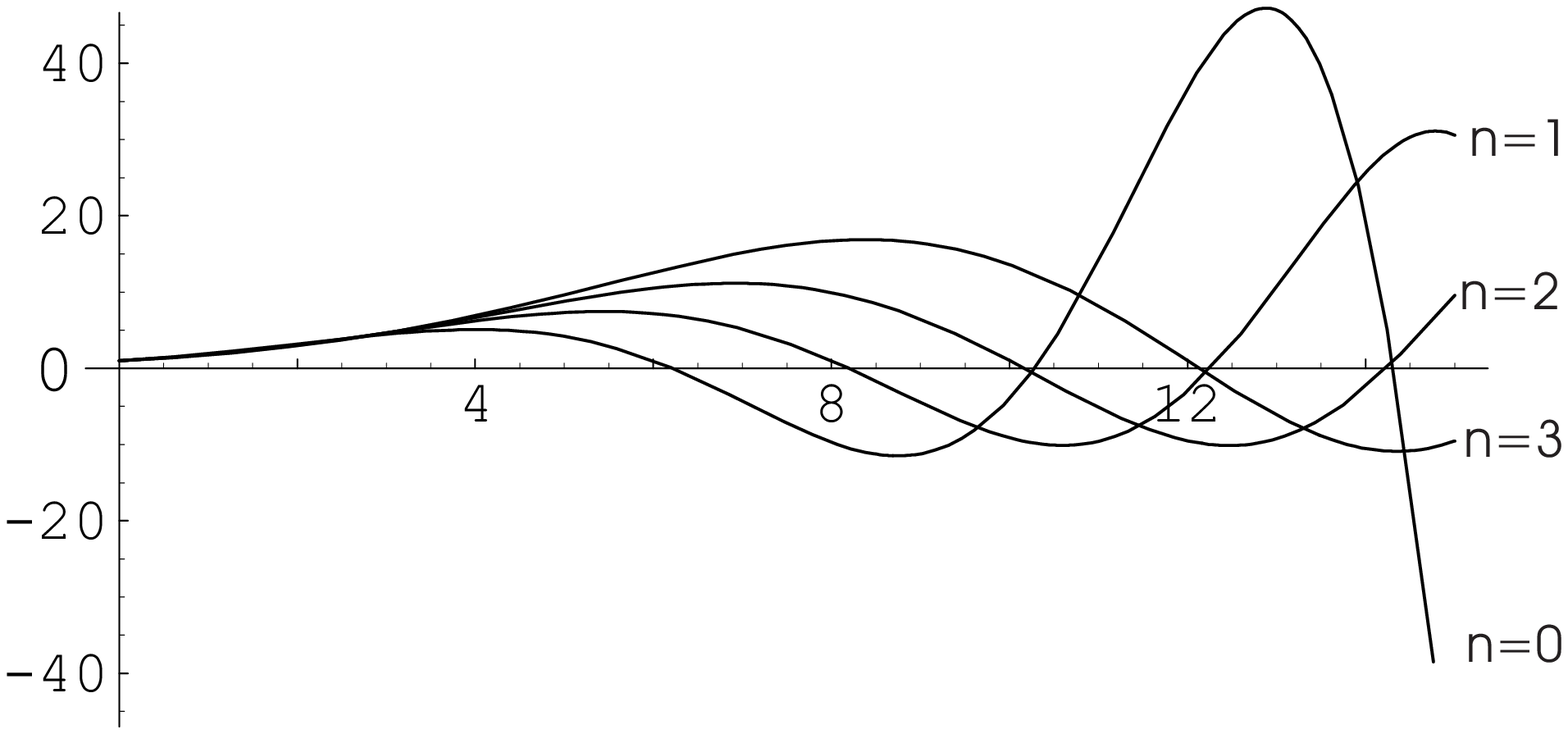}}
\noindent~{\small Fig.~1: Graphical solution of eq.(\ref{f1}) for $n=0,1,2,3$.
The masses for the KK modes of the 4D graviton are at the intersections of the
graphics with the horizontal axis.}
\\

\vspace{.3cm}

\noindent
 {\bf III. Supersymmetry}
\vspace{.2cm}

The next issue we would like to address is supersymmetry. This is
a crucial point as we have assumed that the bulk cosmological
constant is zero and the only proposal we have for achieving this
is bulk supersymmetry. The context here is the chiral 6D
${\cal{N}}=1$ sypergravity.
 We recall that in 6D, the minimal chiral ${\cal{N}}=1$
supersymmetry has vector, hyper, tensor and gravity multiplets. We assume that
the scalars we are employing here belong to a neutral hyper.
 The latter contains, besides $\tau$, others scalar $\{\sigma\}$ and
both $(\tau,\, {\sigma})=\Phi^i, i=1,...,4$ parametrize  the
quaternionic  space $\frac{Sp(n,2)}{Sp(n)\!\times\! Sp(2)'}$.
Supersymmetry is preserved for those backgrounds the gravitino
$\psi_M^A,\, A=1,2$ and hyperino $\psi^a,\, a=1,2$ shifts \be
\delta \psi_M^A=\d_M\epsilon^A+
\frac{1}{2}\omega_{MIJ}\Gamma^{IJ}\epsilon^A+\d_M\Phi^i
Q_i^{AB}\epsilon_B \, , ~~~~~~\delta \psi^a={V_i}^{a A}\epsilon_A
\Gamma^M\d_M\psi^i \ee have zero mode. We have defined
$Q_i^{AB},\, {V_i}^{a A}$ as the composite $Sp(2)'$ connection and
the coset vielbeins, respectively.

It is more convenient, and in order to directly compare with the
 teardrop case, to use the complex scalar $\Phi=(i+\tau)/(1+i \tau)$, employed in
 \cite{Gell-Mann1}, which parametrize $SU(1,1)/U(1)$. In the vacuum we assume that
all  fields except gravity and $\Phi$,  are zero.
In this case, the supersymmetry condition for the vanishing of the
fermionic shifts takes the form
\be
\Gamma^M
\partial_M\Phi\epsilon^A=0\, , ~~~~~~ (\d_M+
\frac{1}{4}\omega_{MIJ}\Gamma^{IJ}-\frac{i}{2}Q_M)\epsilon^A=0
\label{suu}
\ee
where, as usual, $\Gamma^{IJ}=[\Gamma^I,\Gamma^J]/2$ and $Q_M$ is the composite $U(1)$ connection
\be
Q_M=\frac{1}{2 i}\frac{\Phi\d_M\bar{\Phi}-\bar{\Phi}\d_M\Phi}{1-\Phi\bar{\Phi}}
\ee
It is not difficult to see that as long as
the 4D supersymmetry is unbroken, i.e., $\rho_0=0$, the bulk is
also supersymmetric as there are zero modes to eq.(\ref{suu}). This is the teardrop case
and it is known that it preserves susy \cite{Gell-Mann1}.
Indeed, defining $\epsilon^A_{\pm}$ as
$$\Gamma_{\bar{z}}\epsilon^A_+=0,\, \Gamma_z\epsilon_-^A=0,$$ and
using eqs.(\ref{ttt},\ref{2d}) we easily verify that the
background preserve half of the supersymmetries (the one which are
generated either by $\epsilon_+$ or $\epsilon_-$) as the composite
gauge field $Q_M$ exactly cancels the spin
connection \cite{Forste}.

When the 4D supersymmetry is broken, $\rho_0\neq 0$,
there are no solutions to eq.(\ref{suu}). In fact, there exists a solution $\epsilon_+\sim
\bar{z}^{-\alpha}\epsilon_0$, which, however, is not globally defined due to a phase acquired by $\epsilon_+$ as it  goes around $z=0$.
This is a known effect, spaces with deficit angle
do not admit global  Killing spinors~\cite{Henneaux}--\cite{Dvali}.
The only way to keep supersymmetry is to make use of an Aharonov-Bohm phase which will cancel the phase associated to the spin connection
of the conical space \cite{Becker}. For this we need a $U(1)$ vector $A_M$ in the 6D theory which is coupled to the
  gravitino of charge $g$. This means that this $U(1)$ can be a subgroup of the $Sp(2)$ R-symmetry of the
  ${\cal{N}}=1$ 6D theory. Then, we should
replace the derivative acting on $\epsilon$ with $\d_M\epsilon -i gA_M \epsilon$.  The
field strength of $A_M$ should be zero since otherwise it would contribute to the Einstein equations invalidating our solution.
The potential $A_M$ can be non-zero as the conical space is non-simply connected and we may take
$A=A_Mdx^M=
i\beta/2(d z/z-d\bar{z}/\bar{z})$. Clearly, $A$ has vanishing field-strengh $F=dA$ but it cannot gauged away globally
due to the conical structure.
Then the condition for supersymmetry is written as
\be
\d_z\epsilon-ig (i\beta)\frac{1}{2z}\epsilon=0\, , ~~~
\d_{\bar{z}}\epsilon +\frac{a}{\bar{z}} \epsilon-ig (-i\beta)\frac{1}{2\bar{z}}\epsilon=0
\ee
where $a$ is given in eq.(\ref{aaa}).
There is a solution for $\beta=a/2g$, since in this case the Aharonov-Bohm phase cancels the phase acquired by the Killing spinor
going around the apex of the cone at $z=0$.
Thus, in this case supersymmetry is preserved even after brane supersymmetry breaking. An interesting aspect is that althoug
bulk supersymmetry is partially broken, there is no associated Nambu-Golstone fermions as an infrared divergence renders the would-be
Nambu-Golstone fermions  non-normalizable and thus, project them out from the physical Hilbert space \cite{Witten},\cite{Becker},\cite{Forste}.

However,  it is not obvious how you can keep hypers neutral when
the R-symmetry is gauged\footnote{I would like to thank S.
Randjbar-Daemi for stretching out this to me} so the
implementation of the above proposal is not clear. Nevertheless,
the necessary Aharonov-Bohm phase may be provided by the composite
$U(1)$ connection itself. This can be done in the context of the
chiral 6D ${\cal{N}}=1$ or ${\cal{N}}=2$ supergravity. The latter
 can be obtained by compactifying 10D type IIB theory on a
$K3$ surface.
 After an $SU(1,1)$ transformation
$\Phi\to (\rm{u}\Phi+\rm{v})/(\rm{\bar{v}} \Phi+\rm{\bar{u}})$ with $\rm{u}\rm{\bar{u}}-\rm{v}\rm{\bar{v}}=1$, $Q_M$
undergoes the $U(1)$ transformation
$Q_M\to Q_M+\frac{1}{2}\d_M\ln\frac{\rm{u}+\rm{v}\bar{\Phi}}{\rm{\bar{u}}+\rm{\bar{v}}\Phi}$.
By employing
 this observation, one may
 easily verify that the solution
\be ds_2^2={\Big{|}\frac{z}{z_0}\Big{|}}^{-\frac{\rho_0}{\pi
M_*^4}} \left(1-\Big{|}\frac{z}{R}\Big{|}^{\frac{\rho_0}{\pi
M_*^4} }\right)dz d\bar{z} \,
~~~~~~~~\Phi(z)=z^{-\frac{\rho_0}{2\pi M_*^4}} \label{2d} \, .
\label{ssu} \ee indeed satisfies (\ref{suu}) for $\Gamma_{z
\bar{z}}\epsilon^A=\epsilon^A$ and thus, it is supersymmetric.
Note that $\Phi(z)$ is single valued up to an $SU(1,1)$
transformation (with $\rm{u}=\exp(i\frac{\rho_0}{2 M_*^4}),
~\rm{v}=0)$.
\\

\vspace{.3cm}

\noindent
 {\bf IV. Nearby Curved Solutions}

\vspace{.2cm}

We have not examine yet the possibility of curved $4D$ ``lognitudinal" brane metrics in eq.(\ref{MM}).
For  example, we might have looked for solutions of the form
\be
ds^2= e^{2 B(z,\bz)}\hat{g}_{\mu\nu}dx^\mu dx^\nu+e^{2A(z,\bz)}dzd\bz \label{mmm}
\ee
where,
$\hat{g}_{\mu\nu}$ is a metric of de Sitter or anti-de Sitter space of cosmological constant $\lambda$, i.e,
\be
\hat{g}_{\mu\nu}=(-1,e^{2 \lambda t},e^{2 \lambda t},e^{2 \lambda t})\, ,~~~~ \mbox{or}\, ,~~~~
\hat{g}_{\mu\nu}=(-e^{2 \lambda x^3 },e^{2 \lambda x^3},e^{2 \lambda x^3},1)\, , \label{ads}
\ee
respectively.
Although, we cannot prove in general that
there are no such solutions, we can show that flat Minkowski three-branes are the only possibility for
some generic cases like  holomorphic  and  rotational invariant, in the transverse $2D$ space, scalar $\tau$.

It is easy to see that for a $4D$ metric of the form (\ref{ads}), the scalar field equations can  still be solved
for holomorphic $\tau=\tau(z)$. From Einstein equations (\ref{ein2}), we then
get $ B=0\, , \lambda=0$ and, thus, there are no nearby curved solutions to the holomorphic solution.

However, this does not exclude curved $4D$ metrics for non-holomorphic scalar $\tau$.
For example, we may  consider a $6D$ spacetime,  which is maximally symmetric in the  $4D$
lognitudinal brane directions and rotational invariant in the transverse $2D$ space. The symmetry of such a
spacetime is  $G\!\times\!O(2)$,
where $G$ is the Poincar\'e, the de Sitter or the anti-de Sitter group.
We will look for solutions of the form  $A=A(r)\, , B=B(r)\, , \tau=\tau(r)$, where $r^2=z\bar{z}$ such that, for $r$ close to zero
\be
\tau = {\rm const}+\ldots ,~~~ B= - a \log r+\ldots
\ee
as $\tau$ is not coupled to the three-brane and, thus nothing special is experienced by $\tau$ at $r=0$.
It can be then verified that the  Einstein equations (\ref{ein2}) leads to the following behavior of the $A$
\be
A(r)={\rm const}+\ldots\, , ~~~\lambda=0\, .
\ee
As a result,  regularity of $\tau$ at $r=0$, determines the symmetry of the metric to be Poincar\'e$\times O(2)$
 and there are no nearby curved solutions for this case either.
\\

\vspace{.3cm}

\noindent
 {\bf V. Effective $4D$ description and Weinberg no-go theorem}

\vspace{.2cm}

An important aspect is the effective $4D$ theory as it is experienced by a $4D$ observer on the three-brane.
It can be obtained by integrating out
all bulk degrees of freedom. For this, we  compactify the $6D$ theory as described by the action eq.(\ref{action}),
on the internal $2D$ space
with metric {\ref{2d}). The $6D$ metric may then be written as
\be
ds^2=g_{MN}dx^Mdx^N=e^{\Phi}g_{\mu\nu}(x^\mu)dx^\mu dx^\nu+e^{-\Phi}h_{ij}dx^idx^j \label{4d}
\ee
where,
\be
h_{ij}dx^idx^j=
{\Big{|}\frac{z}{z_0}\Big{|}}^{-\frac{\rho_0}{\pi M_*^4}}\left(1-\Big{|}\frac{z}{R}\Big{|}^{2b}\right)dz d\bar{z}
\ee
is the metric of the  internal $2D$ space and its volume has been promoted to a $4D$ scalar $\Phi=\Phi(x^\mu)$. Thus, we are expecting in the
$4D$ effective action a graviton $g_{\mu\nu}$ and a scalar $\Phi$, whereas the graviphoton $A_\mu=g_{\mu\varphi}$, corresponding to the $U(1)$
symmetry of the internal space $z\to e^{i\psi}z$ has been omitted as it is not important for the following. A more complete discussion for the $4D$
spectrum can be found in \cite{Gell-Mann2}.
Using (\ref{4d}) in eq.(\ref{action}), the $4D$ effective action turns out to be
\be
S_{\rm eff}=\int d^4x \sqrt{-g}\left(M_P^2\Big{(}\frac{1}{2} R(g)-\frac{1}{2}\d_\mu\Phi\d^\mu\Phi\Big{)}+e^{2\Phi} c+e^{2\Phi}{\cal{L}}_{SM}\right)
\ee
where
\be
c=\frac{1}{2}\int d^2x \sqrt{h}\, M_*^4\left(R(h)-\frac{1}{2}\frac{h^{ij}\d_i\tau\d_j\bar{\tau}}{\tau_2^2}\right)\label{c}
\ee
The $4D$ Planck mass $M_P$ is finite and given  by
\be
M_P^2=M_*^4 \int d^2x \sqrt{h}=2 \pi \, M_*^4\,\Big{|}\frac{z_0}{R}\Big{|}^{-2a}\frac{b\,  R^2}{ (1-a)(1+b-a)}
\ee
and ${\cal{L}}_{SM}={\cal{L}}_{SM}(e^{\Phi}g,H)$ is the SM action, where by $H$ we denote collectively all SM fields. Integrating out all SM degrees
of freedom and replacing ${\cal{L}}_{SM}$ by the full 1PI effective action $\Gamma_{\rm eff}$, we have in the vacuum
\be
S_{\rm eff}=\int d^4x \sqrt{-g}M_P^2\left(\frac{1}{2} R(g)-\frac{1}{2}\d_\mu\Phi\d^\mu\Phi+e^{2\Phi}C \right) \label{4de}
\ee
where
\be
C=\frac{1}{M_P^2}\left(c-\!<\!\!V_{\rm eff}\!\!> \right) \label{C}
\ee
Finding flat space solution ammounts to minimize the $4D$ effective action (\ref{4de}) for constant metric and scalar $g_{\mu\nu}=\eta_{\mu\nu}\, ,
\Phi-=\Phi_0$. In this case we have
\be
S_{\rm eff}=\int d^4x \sqrt{-g}\, M_P \, C\, e^{2\Phi_0} \label{4dee}
\ee
and we are facing Weinberg theorem: $GL(4)$ invariance determines the form of (\ref{4dee}) and fine-tuning is needed to arrange $C=0$ in order
to satisfy the  field equations. This is
indeed true in a pure $4D$ setup. Here, however, the value of $C$ is determined dynamically as it is related to the background bulk equations.
%
A close insection of eq.(\ref{ein2}), gives then that
\be
 \int d^2 x\, \sqrt{h} \left(R-\frac{1}{2}\frac{h^{ij}\d_i \tau\d_j\bar{\tau}}{\tau_2^2}\right)\Big{|}_{\rm vac.}= 2 M_*^4 \rho_0
\ee
Recalling that $\rho_0=<\!\!V_{\rm eff}\!\!>$ and using eqs.(\ref{c},\ref{C}) we get immediately that
\be
C=0
\ee
Thus, for the $4D$ observer at the three-brane,  the $4D$ cosmological constant is always zero.
As we have only use equations of motions, no fine-tuning required for this result. Weinberg theorem is
valid and the value of the effective action is proportional to the volume element. Just it happens the
proportionality factor to be zero, not because of fine-tuning, but rather due to  bulk dynamics, which forces  the
$4D$ cosmological constant to be zero as a result of  an exact
cancellation of the brane tension by the transverse curvature. This can also be verified by calculating the effective vacuum energy density
$\rho_{\rm eff}$.
The later is given by
\be
\rho_{\rm eff}=\rho_0-\frac{1}{2} \int d^2 z\, e^{2 A}M_*^4 \left(R-\frac{1}{2}\frac{\d \tau\bar{\d}\bar{\tau}}{\tau_2^2}\right)
\ee
and one  easily finds that $\rho_{\rm eff}=0$.
\vskip.2in

We have presented above a partial solution to the cosmological constant problem based on a 6D setup. In particular,
we have shown that a three-brane floating in a six-dimensional bulk is always flat independently of its tension.
Of course this is not enough as there is a number of constraints that should be satisfied, such as
 the finite value of the 4D Planck
scale, the  vanishing of the bulk cosmological constant e.t.c. We have discuss all these constraints in the case of
a finite-volume and non-compact internal space like the tear-drop with a cone singularity. It seems that
the solution we propose is consistent with conservation laws, KK spectrum, supersymmetry,
nearby curved configurations as well as with Weinberg theorem.
We should stress that there are no fine-tunings involved in the present setup. The cone is adjusted itself
for any value of the cosmological constant
such that the biggest the cosmological constant, the sharpest the apex of the cone and the largest the volume of the tear-drop.
In a sense, the cone acts as a drain, depleting vacuum energy from the brane to the tear-drop increasing its volume.

\[ \]
{\bf Acknowlegements} We wish to thank S. Dimopoulos and T.
Gherghetta for intensive discussions. This work is partially
supported by the EPAN projects, Heraclitus and Pythagoras,
 and the 8198 Protagoras NTUA project.


\end{document}